\documentclass[aps,prl,twocolumn,showpacs]{revtex4}
\usepackage[dvips]{graphicx}
\usepackage{psfig,epsfig}
\usepackage{amsmath}
\usepackage{amsfonts}
\usepackage{amssymb}
\usepackage{wasysym}

\def\be{\begin{equation}}
\def\ee{\end{equation}}
\def\bfi{\begin{figure}}
\def\efi{\end{figure}}
\def\bea{\begin{eqnarray}}
\def\eea{\end{eqnarray}}
\begin{document}
\voffset=+1truecm
\title{Universality in solar flare and earthquake occurrence}
\author{L. de Arcangelis$^{a}$, C. Godano$^{b}$,
E. Lippiello$^{c,}$}\altaffiliation { present address: 
Department of Physics ``E.R. Caianiello'', University of Salerno,
 84081 Baronissi (SA), Italy}
\author{
 M. Nicodemi$^{d}$} 
\affiliation{ $^{a}$ 
Department of Information Engineering and CNISM, 
Second University of Naples, 81031 Aversa (CE), Italy\\ 
$^{b}$ 
Department of Environmental Sciences and CNISM, Second University of
Naples, 81100 Caserta, Italy\\
$^{c}$ University of Naples ``Federico II'', 80125 Napoli, Italy\\ 
$^{d}$ 
Department of Physical Sciences,
University of Naples ``Federico II'', Coherentia-CNR and INFN,
80125 Napoli, Italy }

\begin{abstract}
Earthquakes and solar flares are phenomena involving huge and rapid 
releases of energy characterized by complex temporal occurrence. 
By analysing available experimental catalogs,
we show that the stochastic processes underlying these apparently different
phenomena have universal properties. Namely both problems 
exhibit the same distributions of 
sizes, inter-occurrence times and the same temporal clustering:
we find afterflare sequences with power law temporal correlations as the 
Omori law for seismic sequences.
The observed universality suggests a common approach to 
the interpretation of both phenomena in terms of the same 
driving physical mechanism. 
\end{abstract}


\maketitle

Solar flares are highly energetic explosions \cite{Phillips} 
from active regions of the Sun in the form of electromagnetic radiation, 
particle and plasma flows powered by strong and twisted magnetic fields.
Since they cause
disturbances on radio-signals, satellites and electric-power on the Earth,
much interest has been devoted in the
last years to space weather forecasts \cite{spaceweather,spaceweather2}. 
Recent studies have shown that solar flares also affects the Sun's interior, 
generating seismic waves similar to earthquakes \cite{koso}. Actually, 
despite having different causes, solar
flares are similar to earthquakes in many respects, for example
in the impulsive localised release of energy and momentum and their
huge fluctuations \cite{laza}.
The analogy with earthquake occurrence is also
supported by the observation of power laws \cite{McTiernan,Kane,Crosby}
in the distribution of flare sizes, $P(s)$,
related to the Gutenberg-Richter law for the earthquake magnitude
distribution. 
Various interpretations have been proposed for these power law distributions 
ranging from Magneto-Hydro-Dynamics \cite{Phillips,MHD} to turbulence
\cite{Boffetta} up to Self Organized Criticality \cite{Lu,HNJ,Hug}.
A better understanding of solar flares and coronal mass ejections from
the Sun requires knowledge of the structural details of these events and 
their occurrence in time.
This could greatly improve the prediction of violent space weather
and the understanding of the physical processes behind solar events.

Here we present evidence that the same empirical laws widely accepted in 
seismology characterize, surprisingly, also the size and time occurrence 
of solar flares. 
In particular the same temporal clustering holds both for earthquake, 
where it is known as the Omori law, and solar flare catalogues: 
a mainflare triggers a sequence of afterflares. 
The evidence of a universal statistical behaviour suggests the possibility 
of a common approach to long term forecasting and rises as well deep 
questions concerning the nature of the common basic mechanism.

A statistical approach to earthquake catalogues has revealed a scale invariant 
feature of the phenomenon, as indicated by power law
distributions for relevant physical observables \cite{Knopoff,BT}, 
such as the seismic moment distribution of earthquakes, 
$P(s) \sim s^{-\alpha}$, where the exponent $\alpha \in [1.6,1.7]$ is 
essentially the same in different areas of the world \cite{Kagan}. 
This relation corresponds to the Gutenberg-Richter law for the distribution 
of the earthquake magnitude $M$ via the relation 
$M = 2/3\log(s)-K$, where $K$ is a constant \cite{Kanamori}. 
It is also observed 
that earthquakes tend to occur in clusters temporally located after 
large events: the Omori law states that the number of aftershocks 
at time $t$ after a main event, $N_{A}(t)$, decays 
as a power law $N_{A}(t) \sim t^{-p}$ with $p \simeq 1$ \cite{Omori}. 
Finally, the distribution of intertimes between consecutive earthquakes, 
$P(\Delta t)$, 
is not a simple power law, but has a non trivial functional form which, 
like the other quantities mentioned before, is essentially independent 
of the geographical region or the magnitude range considered \cite{Corral}. 
These observations suggest that $P(\Delta t)$, $N_{A}(t)$ and $P(s)$ are 
distinctive features of earthquakes and, thus, fundamental quantities for 
a probabilistic analysis of the phenomenon characterizing its amplitude and 
time scales.

In this letter we analyse several catalogues of solar flares and compare 
them with the Southern California catalogue for earthquakes.
Since emissions at different
wavelengths are related to different radiation mechanisms, we
present a comparison among solar data  from X-ray observations in
three different energy ranges and different periods of solar
cycle, by using on-line available flare catalogues. More
specifically, Soft X-ray data in the (1.5-2.4) keV and (3.1-24.8)
keV ranges are obtained from the Geostationary Operational
Environmental Satellite (GOES) systems \cite{GOES}. We consider
$N_e=21567$ events from January 1992 to December 2002 covering the
entire 11-years solar cycle. Solar flares in the Hard X-ray range
($>$25 keV) are obtained from the Burst and Transient Source
Experiment (BATSE) that gives $N_e=6658$ flares from April 1991 to
May 2000 \cite{BATSE}. Finally $N_e=1551$ events from January 1990
to July 1992 in the Intermediate X-ray (10-30 keV) range are
analysed from the WATCH experiment \cite{WATCH}. 
Many earthquake catalogues exist and the universality of their 
statistical features has already been established \cite{Corral,Kagan}. 
Thus, for clarity, we only consider here the Southern California 
earthquake catalogue \cite{California} which has $N_e=88470$ events 
with magnitude $M\ge 2$ in the years from $1967$ to $2002$. 

\begin{figure}
\hspace{-2cm}
\includegraphics[width=9cm,angle=-90]{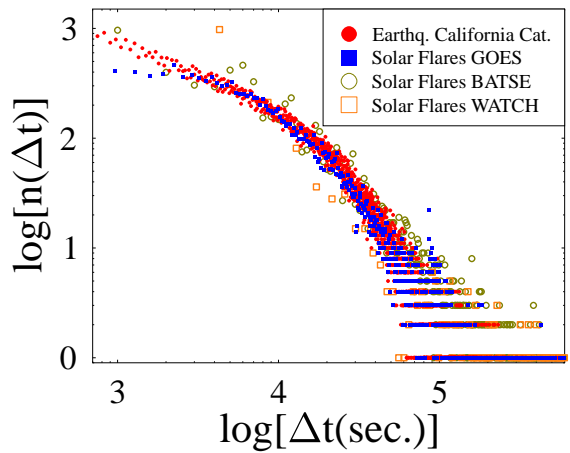}
\vspace{-1cm}
\caption{(Color online)
The number distribution, $n(\Delta t)$, of intertimes, $\Delta t$,
between consecutive events in solar flare and earthquake catalogues.
Solar data refer to X-ray observations in three different energy
ranges covering different periods of solar cycle:
soft X-ray data in the (1.5-2.4) keV and (3.1-24.8) keV ranges
from the GOES catalogue ($\blacksquare$ symbols);
hard X-ray ($>$25 keV) from the BATSE catalogue ({\Large $\circ$} symbols);
intermediate X-ray (10-30 keV) from the WATCH catalogue ($\square$ symbols).
Earthquake intertimes data are from the
California catalogue for events with magnitude $M\geq 2$
({\Large $\bullet$} symbols).
}
\label{fig1}
\end{figure}

The intertime distribution has been already investigated both for 
earthquakes \cite{Corral} and solar flares \cite{whe3}. 
The intertime, $\Delta t$, is the time between the start of a flare 
(or an earthquake) and the start of the next one as reported in the above
catalogues. Here, for a catalogue with $N_e$ data, we count the number of
events $n(\Delta t)$ having intertime between $\Delta t$ and
$\Delta t+\lambda/N_e$, where $\lambda$ is a constant setting the
binning of raw data. This is the
statistically relevant quantity to consider \cite{statistica}, 
since $n(\Delta t)/\lambda \to P(\Delta t)$ in the limit $N_e \to \infty$, and,
thus, in the following we refer to $n(\Delta t)$.
Here we choose $\lambda$ such that $\lambda/N_e=75 sec$ for the California 
catalogue and use the same value for all catalogues.
Fig.1 shows the intertime distributions, $n(\Delta t)$, for the three
different solar flares data sets and for the California
earthquake catalogue. Solar flares data scale
one on top of the other to a very good approximation and,
interestingly, they all appear to collapse, within statistical
errors, on the same non trivial distribution function of earthquake
intertimes. In particular, this data
collapse is obtained without rescaling $\Delta t$ by any suitable
factor: the intertime duration, $\Delta t$, is expressed in the
same units (seconds) for all data sets. Thus, Fig.1 shows that
the same intertime distribution function and,
surprisingly, even the same time range characterize these apparently
different physical processes in the magnitude range of the above catalogues.

\begin{figure}
\hspace{-2cm}
\includegraphics[width=9cm,angle=-90]{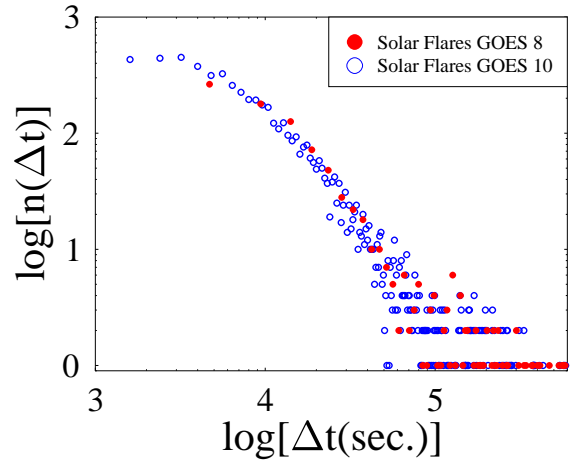}
\vspace{-1cm}
\caption{
(Color online) The number distribution, $n(\Delta t)$, of intertimes, 
$\Delta t$, between consecutive solar flares for the GOES catalog. Data
from GOES8 and GOES10 satellites correspond to the minimum of the solar 
cycle (from 9/1/1995 to 12/31/1996) and the maximum (from 1/1/2000 to
12/31/2003) respectively. 
}
\label{fig2}
\end{figure}

The scaling behaviour of $n(\Delta t)$ for different solar phases is a widely 
debated subject \cite {whe1,whe2,Pacz} and a dependence on the solar phase
\cite{whe1} has been observed also in the case of Coronal Mass Ejection 
\cite{whe2}. 
The result for flares has been obtained by taking into 
account only events with a peak flux larger than $1.4\times 10^{-6}
Wm^{-2}$
 (class C1). We have then considered separately data from the GOES 
catalog corresponding to maximum and minimum solar activity and used the same
binning procedure as Fig.1. In order to take into account the different
level of background X-ray flux, we have set different thresholds for different
phases: events greater of class C1 in the maximum and class B1 ($10^{-7}\times
Wm^{-2}$) in the minimum phase \cite{nota2}.	
Fig.2 shows that data from different phases fall on the same universal curve. 

\begin{figure}
\hspace{-2cm}
\includegraphics[width=9cm,angle=-90]{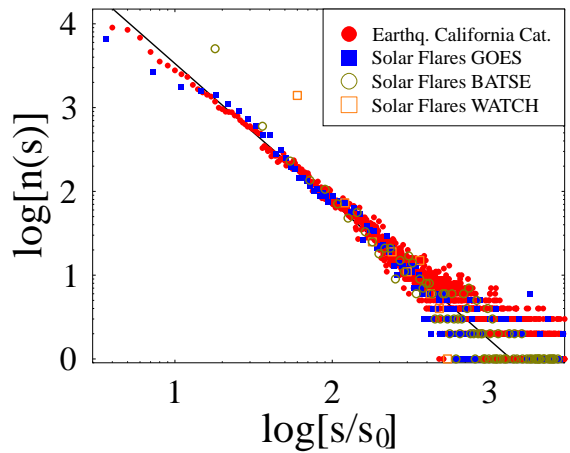}
\vspace{-1cm}
\caption{
(Color online) The distribution, $n(s/s_0)$,
of flare peak intensity, from the same catalogues of
Fig.1, and of seismic moments from the California earthquake catalogue.
We set $\lambda/N_e=1$ for the California catalogue.
As for the intertimes of Fig.1, comparison between the
size distributions of earthquakes and solar flares shows
very good agreement. The universal distribution is
well fitted by a power law with exponent $\alpha =
1.65\pm 0.1$ (shown as a solid line in the picture).
$s_0= 10^{-7}$ W/m$^2$ for GOES, $s_0=600$ cmnts/(sec
2000cm$^2$) for BATSE, $s_0=3000$ cnt/bin for WATCH, $s_0=30\cdot
10^{16}$ Nm for earthquakes.
}
\label{fig3}
\end{figure}

The other crucial quantity to be investigated is the distribution
of flares sizes, $P(s)$, i.e., the distribution of the flare peak
intensity, $s$, from the above catalogues. This is
compared with the earthquake seismic moments distribution, $P(s)$, from
the Southern California catalogue. In order to normalize the
different units and experimental ranges used in each catalogue, here
we scale the values, $s$, of each data set by a given constant amount
$s_{0}$. 
Then, we calculate the number of events $n(s/s_{0})$ with
sizes between $s/s_0$ and $s/s_0+\lambda/N_e$. 
Excellent data collapse is observed in Fig.3 
with all data fitted over almost three decades
by a power law $n(s/s_{0}) \sim (s/s_{0})^{-\alpha}$ with an
exponent $\alpha =1.65 \pm 0.1$, in agreement with previous
results on solar flares \cite{McTiernan,Kane,Crosby} and
earthquakes \cite{Kagan}.
Therefore, in analogy to earthquakes, from the above observations
it is possible to introduce a Richter scale for flares where
their ``magnitude'', $M$, is defined via the relation:
$M(s)=2/3\log(s)-K_F$, where $K_F$ is a  constant.
In terms of the magnitude the data from flares catalogs are therefore
found to follow the Gutenberg Richter law introduced for earthquakes.

Further evidence of structural similarities in the statistics of the 
two phenomena is given by the analysis of correlations between events 
within each of these catalogues. 
It would be interesting to compare the time correlation
between main-events and the sequence of their after-events, as in
the Omori law. 
We define a ``main-event" as an event with magnitude $M>M_{main}$; 
its ``after-events" are the following events with 
$M_{cut}< M<M_{main}$, where $M_{cut}$ is a cutoff for small background events.
The basic difference with the usual definition used in seismology 
is that an event with  
$M<M_{main}$ considered as ``aftershock'' may instead be an  independent
event totally unrelated to the preceding ``mainshock". Furthermore, 
an event with $M>M_{main}$ considered as ``mainshock''  
may have been triggered by a previous 
larger event. Despite these differences, our definition can be 
straightforwardly applied to flare catalogues too and 
tends to the standard one for large enough $M_{main}$ and 
$M_{cut}$: here we fix $M_{cut}= M_{main}-2.5$.
In Fig.4 we show the number of ``after-event", $n_{A}(t)$, at time $t$ 
after a ``main-event", for all the mentioned data sources. 
Interestingly, the time correlation function, $n_{A}(t)$, has the same 
functional form in all catalogues.
A power law $n_{A}(t)\sim 1/t$ (straight line in the picture),
as the Omori law, fits the data. The results are
quite robust with respect to changes of the parameter 
$M_{main}$, provided that $M_{main}$ is large enough as previously explained.
We apply the same procedure to analyse the rate of occurrence of events
leading up to a main event and observe that also "fore-flares" follow the same
power law behaviour (Omori law) as foreshocks \cite {yama}, even if more
symmetrical behaviour is observed in the  flare case.

\begin{figure}
\hspace{-2cm}
\includegraphics[width=9cm,angle=-90]{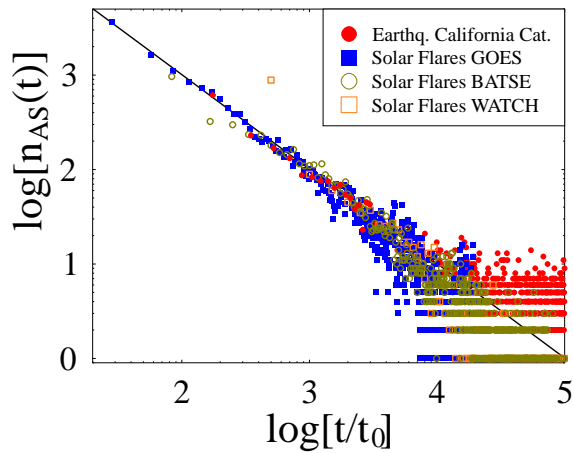}
\vspace{-1cm}
\caption{
(Color online) The correlation function, $n_{A}(t)$, i.e., the number of
``after-event'' at time $t$ after a ``main-event'' for
the same catalogues of Fig.1.
To find the best collapse, data from different catalogues are rescaled by
a given amount $t_0$ ($t_0=700$ sec for GOES, $t_0=60$ sec for BATSE, $t_0=1$
sec WATCH and $t_0=43$ sec for California earthquakes).
As for Fig.1 and Fig.3,
the ``aftershock'' rate of occurrence for earthquakes and solar flares scale
very well. For comparison, we also plot an Omori power law
$n_{A}(t/t_0)\sim t_0/t$ (solid line).
}
\label{fig4}
\end{figure}

Fig.s 1, 3 and 4 indicate that the statistical properties
of size and time scales of solar flares (independently of the energy range
and the temporal location in the solar cycle of the X-ray radiation)
and earthquakes are essentially the same within current statistical accuracy.
It is tempting to look at the observed universality in the perspective of 
the theory of critical phenomena. In the past 
analogy between the two phenomena was proposed on the basis of the same theoretical model
\cite {BTW}. Here we follow a completely different approach: we
directly compare experimental catalogs, we observe universal behaviour
and therefore we propose the presence of a common driving physical mechanism. 
Most earthquakes occur where the elastic energy builds up owing to
relative motion of tectonic plates. Schematically, as the friction locks the
sliding margins of the plates, energy load increases. When elastic stress
overcomes the threshold of frictional resistance, it is relaxed
causing the occurrence of an earthquake. This ``stick-slip" behaviour
redistributes the stress-energy field in the crust generating new 
earthquakes where and when the local slipping threshold is exceeded. 
A quantitative prediction on the aftershocks decay cannot be
derived by simple stress transfer but can be obtained 
in terms of a state variable constitutive formulation, where 
the rate of earthquakes results from the applied stressing 
history \cite{Dieterich}. This formulation gives account for 
long-range correlations between earthquakes affecting
the shape of the whole intertime distribution. 

The universal scaling of Figures 1-4 suggests 
a similar physical mechanism 
at the basis of solar flares occurrence.
Flares and X-ray jets arise in active solar regions where
magnetic flux emerges from the solar interior and interacts with 
ambient magnetic field. These interactions are thought to occur in
electric current sheets separating regions of opposite magnetic
polarity. The dynamics and energetics of these sheets are governed by
a complex magnetic field structure \cite{sol}. 
Opposite fluxes lead to rearrangement of field lines building
up magnetic stress up to a breaking point where magnetic energy is released
in a flare via magnetic reconnections. 
The observed temporal clustering of Fig. 4 shows that the rate of
flare occurrence decreases in time as a power law after a "mainflare".
Since the same behaviour is found for seismic sequences, here we
propose that the mechanism at the basis of seismic energy redistribution 
can be responsible for "afterflare" occurrence. In particular 
magnetic stress transfer in Solar Corona plays the role of elastic stress 
redistribution on the Earth crust. As a consequence the state-rate formulation
\cite{Dieterich} can be generalized to solar flares, namely the flare 
triggering depends on the entire history of magnetic stresses. 
Beyond issues of fundamental science, the present
results can also have very practical consequences as, for instance, 
to improve the prediction of violent space weather by applying 
established methods of seismic forecasting \cite{Reasenberg}. 

{\small Acknowledgments. This work is part of the project of the Regional
Center of Competence ``Analysis and Monitoring of Environmental Risk",
supported by the European Community on Provision 3.16.
}


\begin{thebibliography}{40}
\bibitem{nota2} In order to choose the thresholds, we verify the power law
behaviour of the peak flux distribution for energy above threshold. In
seismicity this criterion is a test for catalog completeness.
\end{thebibliography}
\end{document}